\pdfoutput=1 
\documentclass[aps,groupaddress,superscriptaddress,onecolumn,amsmath,amssymb,graphicx
]{revtex4-2}

\usepackage{graphicx}
\usepackage{amsmath}
\usepackage{color}
\usepackage{float}

\begin{document}

\title{Generalization of Fluctuation-Dissipation Theorem to Systems with Absorbing  States}

\author{Prajwal Padmanabha\footnote{Corresponding author. Email: pprajwal122@gmail.com}}
\affiliation{Laboratory of Interdisciplinary Physics, Department of Physics and Astronomy ``G. Galilei'', University of Padova, Padova, Italy}
\author{Sandro Azaele}
\affiliation{Laboratory of Interdisciplinary Physics, Department of Physics and Astronomy ``G. Galilei'', University of Padova, Padova, Italy}
\affiliation{INFN, Sezione di Padova, via Marzolo 8, Padova, Italy - 35131}
\affiliation{National Biodiversity Future Center, Piazza Marina 61, 90133 Palermo, Italy}
\author{Amos Maritan}
\affiliation{Laboratory of Interdisciplinary Physics, Department of Physics and Astronomy ``G. Galilei'', University of Padova, Padova, Italy}
\affiliation{INFN, Sezione di Padova, via Marzolo 8, Padova, Italy - 35131}
\affiliation{National Biodiversity Future Center, Piazza Marina 61, 90133 Palermo, Italy}

\begin{abstract}
\textbf{Abstract:} Systems that evolve towards a state from which they cannot depart are common in nature. But the fluctuation-dissipation theorem, a fundamental result in statistical mechanics, is mainly restricted to systems near-stationarity. In processes with absorbing states, the total probability decays with time, eventually reaching zero and rendering the predictions from the standard response theory invalid. In this article, we investigate how such processes respond to external perturbations and develop a new theory that extends the framework of the fluctuation-dissipation theorem. We apply our theory to two paradigmatic examples that span vastly different fields - a birth-death process in forest ecosystems and a targeted search on DNA by proteins. These systems can be affected by perturbations which increase their rate of extinction/absorption, even though the average or the variance of population sizes are left unmodified. These effects, which are not captured by the standard response theory, are exactly predicted by our framework. Our theoretical approach is general and applicable to any system with absorbing states. It can unveil important features of the path to extinction masked by standard approaches. \\

\end{abstract}

\maketitle 

\section{Introduction}

Statistical physics has been an invaluable bridge between microscopic dynamics and macroscopic observables. The Einstein-Smoluchowski equation, the regression hypothesis by Onsager \cite{marconi2008fluctuation,onsager1931reciprocal1,onsager1931reciprocal2}, and linear response theory by Kubo \cite{kubo1957statistical,kubo1966fluctuation} are different kinds of fluctuation-dissipation theorems (FDTs).  These relations, which were also extended to stochastic systems \cite{risken1996fokker,zinn1996quantum}, connect the response of an observable to stationary correlations. This allows for the prediction of the effects of perturbation from equilibrium dynamics. Though equilibrium dynamics is a useful description, studies into non-equilibrium systems have gained much traction due to the vast number of examples showing non-equilibrium behaviour. These examples include processes from biology, epidemiology, turbulence, and chemical systems, amongst others \cite{fang2019nonequilibrium,de2013non,kamenev2008extinction,cardy2008non}.

Given the predictive power of these theorems, their violations also become crucial to study. A substantial body of literature exists on the violation of the fluctuation-dissipation theorem in glassy systems \cite{crisanti2003violation, baity2017statics}. In non-glassy systems, there have been attempts at extending the validity of FDT to non-equilibrium dynamics using the concepts of asymmetry \cite{cugliandolo1994off,diezemann2005fluctuation}, frenesy \cite{baiesi2009fluctuations}, and local currents \cite{seifert2010fluctuation,chetrite2008fluctuation,verley2011modified1}. In particular, successes have been obtained at predicting response to perturbation in an arbitrary non-stationary state \cite{verley2011modified2}. However, the prevalent and inherent assumption is that there is some form of eventual non-trivial steady state. Instead, the focus of this work is on an entirely different class of non-equilibrium systems with absorbing states/boundaries.

Processes with absorbing states are ubiquitous in nature. Some examples include chemical reactions, epidemics, and population dynamics \cite{bartlett1960stochastic,azaele2016statistical,gardiner1985handbook,van1992stochastic}. In these cases, the net outward flux of the system's constituents is positive, and hence the total probability distribution decays with time, rendering the steady-state trivial. However, they are also frequently found in a quasi-stationary regime for a long time before reaching extinction.  There exists extensive mathematical literature analyzing the properties of quasi-stationary systems \cite{pollett2008quasi,meleard2012quasi,van1992stochastic}. The details of time taken to reach extinction, the existence of quasi-stationary distribution \cite{darroch1965quasi,darroch1967quasi}, simulation methods \cite{de2005simulate}, etc have been studied with applications to cellular automata \cite{atman2002quasistationary}, birth-death processes \cite{van1991quasi}, Brownian motion \cite{martinez1994quasi}, contact process \cite{dickman2005quasi}, and many others. 

Despite the vast literature, there have been no forays through the lens of statistical physics into looking at the change in FDT in the presence of an absorbing state. There is a gap in our understanding of the effects of perturbation in decaying processes. Linear response theory in ``quasi-stationary'' states has been previously considered \cite{ogawa2012linear,patelli2012linear}, but the states referred to are metastable states in which systems spend a long time before reaching equilibrium, rather than processes reaching extinction. In general, fields of quasi-stationary systems and non-equilibrium FDT are both crucial and significant, but they have not been connected.  

In the following work, we build the bridge between these two fields by considering the modifications needed by linear response theory to accommodate the presence of absorbing states. We consider the method of conditioning observable averages over trajectories that have not yet hit the absorbing state, calling this operation conditioning to survival. With this framework, we find that a generalized FDT arises by introducing a new observable that accounts for the survival of trajectories. This theorem is a true generalization and cannot be obtained by any ad hoc replacement of the stationary distribution in the standard case.

To demonstrate the scope of the approach, we illustrate applications in two different and wide-ranging contexts (ecological and biochemical), both having important practical implications.

One of the standard techniques used for stochastic modelling of forests is the birth-and-death process \cite{karlin1958linear,azaele2015towards}. This one-step Markov process, when considered with constant per capita birth rate ($b$), death rate ($d$) and immigration, results in the well-known Fisher log-series stationary distribution \cite{volkov2003neutral} which gives the relative species abundances (RSA) and a biodiversity parameter in an ecosystem. Under the absence of immigration, including the zero population absorbing state (and considering $b<d$), we show the validity and importance of our theory, comparing it to the equivalent of the standard case, where the response is significantly underestimated. We then proceed to show that, while common observables like population size average and variance, and even the observed RSA can remain constant under perturbation, the extinction probability for a species can irreversibly be changed.

On the other scale, we pick the example of targeted search by proteins on DNA. This process is modelled using hybrid diffusion models where the protein moves along the DNA searching for a specific set of base pairs (1D diffusion) but can also detach temporarily and move in free space (3D diffusion) \cite{von1989facilitated,berg1981diffusion} leading to speeding up of the target finding rate \cite{halford2004site,kolomeisky2011physics,kamagata2020p53}. In this example, we consider a discrete state version of 1D diffusion of the protein, with a possible detachment to 3D space and reattachment to the DNA. This kind of discrete model has been used previously to explain different properties of the process, especially optimal target-finding speed \cite{shvets2018mechanisms,iwahara2021discrete}. Here, we predict the distance of the protein from the target under changing temperature conditions and proceed to show how the distribution of absorption time skews towards faster absorption with a sudden perturbation in the detachment rate.  

In both examples, we compare analytical predictions to numerical simulations of the examples and find excellent agreement between the two. 

\section{Generalization of fluctuation-dissipation theorem} 

We begin with a recap of the standard FDT that we will later use to compare the generalized version. Consider a stochastic system comprised of a finite set of discrete states, represented by $\Omega$. Let $P_x(t)$ be the probability of finding the system in state $x \in \Omega$ at time $t$. The evolution of the probability $P_x(t)$ is governed by the Master equation
\begin{equation}
    \dot{P}_x(t) = \sum_{y\neq x} \big[ W_{xy}~P_y(t) - W_{yx}~P_x(t)\big]
\end{equation}
where $W_{xy}$ is the rate of transition from state $y$ to state $x$. By defining a matrix with elements $H_{xy} \equiv W_{xy} - \delta_{xy} \sum_z W_{zy}$, the Master equation can be written as 
\begin{equation}
    \dot P(t) = H~P
    \label{def:master-equation}
\end{equation}
and hence has the solution $P(t) = P(0)~e^{H~t}$ \cite{van1992stochastic}. Without any absorbing boundaries, at large times, if the process is irreducible and positive recurrent \cite{durrett2019probability}, a stationary state is reached, where the probability of finding the system in state $x$ is $P^{st}_x$. A perturbation is described by $H+\delta H(t)$, with $\delta H(t)= F(t)\delta H$ where $F(t)$ is  the time component of $\delta H$. If the transition rate matrix $H$ depends on a certain set of parameters $f \equiv (f_1, f_2 , \dots)$, the perturbation changes at least one of them, resulting in $f \to f + \Delta f(t)$. Then, $\delta H(t) = (\partial_f H) ~ \Delta f(t)$ at leading order and the response of any observable $A$ is given by the derivation from its average stationary value according to \cite{risken1996fokker}
\begin{align}
\langle \Delta A(t) \rangle &= \int_{-\infty}^{\infty}dt' R_{A,f}(t-t') \Delta f(t') \label{def:response-function-main} \\
R_{A,f}(t) &=  \Theta (t) \sum_{x,y,z} A_x~[e^{Ht}]_{xy}~(\partial_f H)_{yz}~P^{st}_z \label{eqn:risken-response} \\ 
&= -\Theta(t) \dfrac{\partial}{\partial t} \bigg \langle A(t)~\dfrac{\partial \phi}{\partial f}(0) \bigg \rangle^{st}\label{eqn:fdt-equilibrium}
\end{align}
where $\langle \cdots \rangle $ represents the average over the probability distribution, $R_{A,f}(t)$ is the response function which determines the response of $A$ to perturbation $f + \Delta f(t)$, $\Theta (t)$ is the Heaviside step function, and $\phi_x\equiv\ln P^{st}_x$ is the potential characterising the stationary solution (See Supplementary Material Section 1 for detailed derivation). This equation connects the fluctuation of two variables at equilibrium to the dissipation back to equilibrium after a perturbation by expressing the response function as a time derivative of a two-time correlation function in unperturbed dynamics. Hence, such a relation is termed the fluctuation-dissipation theorem.

In case the process has absorbing states, they are characterised by outgoing transition rates being zero in the master equation, i.e., $W_{x,y} = 0 \quad \forall x \in \Omega \;, y \in  \partial \Omega$, where $\partial \Omega$ the set of absorbing states. We are interested in how the observables can be measured in the interior $\Omega^{\circ}=\Omega -\partial \Omega$. We collapse all the absorbing states into a single representative state $o$ to avoid degeneracy. We are then able to write the master equation in the same form as Eq. \eqref{def:master-equation}, but with redefined transition rates on $\Omega' = \Omega^{\circ} \cup \{o\}$. The right eigenvectors and eigenvalues of $H$ will be denoted as $\varphi_n$ and $\lambda_n$. The entries of the first column of $H$ are all zeros; hence the first eigenvalue is $\lambda_0 = 0$ and the corresponding first right eigenvector is $\varphi_{0,x} = \delta_{x,o}$. This corresponds to the long-time limit of the probability distribution, which is zero everywhere on the interior. We consider the matrix $W$ to be an irreducible matrix, which means there always exists a non-trivial path from any interior state to any other interior state. From the Perron Frobenius Theorem \cite{seneta2006non}, the eigenvalue spectrum is
\begin{equation}
0 = \lambda_0 < \Re(\lambda_1) < \Re (\lambda_2) \leq \Re (\lambda_3) ... 
\label{assum:eigenvalue}
\end{equation}
where $\Re(\cdot)$ denotes the real part of the argument. The second eigenvalue $\lambda_1$ is real and the corresponding left and right eigenvectors are real with positive components in the interior 
(See Supplementary Material Section 2.1.1 for proof and details about the heirarchy. See also \cite{monthus2021large,monthus2022conditioned}). 

Then, the propagator giving the probability of being in the state $x$ at time $t$ starting from $y$ at time $t'$ can be written in terms of the eigenvectors and eigenvalues of $H$, i.e., $P(x,t|x_0,0) = \sum_n c_{n,x_0}~\varphi_{n,x}~e^{-\lambda_n t}$ where $c_{n,x_0}$ are the constants depending on the initial condition. For times $t$ such that $t(\Re (\lambda_2) - \lambda_1) >> 1$, the probability distribution is approximated by $P^{lt}_x(t) = \varphi_{0x} + c_1~\varphi_{1x}~e^{-\lambda_1 t}$.

To ensure that the system has not reached the boundary yet, we introduce an observable that characterizes the survival, namely, $\chi_x = 1 \;\; \forall x \in \Omega^{\circ}$ and $\chi_o = 0$. The conditional average of an observable $A$ on the interior (i.e., $A_{o} = 0$) reaches a stationary  value as $t \to \infty$, i.e.,
\begin{align}
&\langle A || \chi \rangle_t \equiv \dfrac{\langle A(t) \rangle}{\langle \chi(t) \rangle} = \dfrac{\sum_x A_x P_x(t)}{\sum_x \chi_x P_x(t)} = \dfrac{\sum_{x \neq o} A_x P_x(t)}{\sum_{x \neq o} P_x(t)} \nonumber \\
&\lim_{t \to \infty} \langle A || \chi \rangle_t = \lim_{t \to \infty}\dfrac{\sum_{x \neq o} A_x \sum_i c_i~\varphi_{i,x}~e^{-\lambda_i t}}{\sum_{x \neq o} \sum_i c_i~\varphi_{i,x}~e^{-\lambda_i t}}= \sum_{x \neq o} A_x \varphi_{1x}  \equiv \langle A  || \chi \rangle^{lt}
\label{eqn:stationary-conditional-average}
\end{align}
where we have used the eigenvector expansion for the propagator and $\varphi_{1x}$ is the eigenvector associated with $\lambda_1$ (See Supplementary Section 2.1.2 for more details). It is clear from this definition that we are conditioning the observable to survive on the interior. The superscript $lt$ is used to denote the conditional observable average at large times. Similarly, we write the conditional correlation between two observables $A$ and $B$ at times $t$ and $t'$, with $t > t'$, which at sufficiently large starting times gives
\begin{align}
&\langle A(t)B(t') || \chi (t) \rangle = \dfrac{\langle A(t)B(t')\rangle}{\langle\chi(t)\rangle} = \dfrac{\sum_{x,y} A_x ~P(x,t|y,t')~B_y P_y(t')}{\sum_x \chi_x P_x(t)}  \nonumber \\
&\lim_{t,t' \to \infty} \langle A(t)B(t') || \chi (t) \rangle =  \sum_{x,y} A_x~B_y~P(x,t|y,t')~\varphi_{1y}~e^{\lambda_1 (t-t')} \equiv \langle A(t)B(t') || \chi \rangle^{lt} 
\label{eqn:stationary-conditional-correlation}
\end{align}
The averages in Eq. (\ref{eqn:stationary-conditional-average}) and Eq. (\ref{eqn:stationary-conditional-correlation}) are equal to averaging the observables directly over $P^{lt}_x(t)$ which is self-consistent  since both are considered at long times. 

A perturbation in the parameters can be represented by a perturbation in the generator $H \to H + \delta H (t)$, leading to a change in the long-time probability distribution, $P(t) = P^{lt}(t) + \delta P(t)$. The solution from the first order differential equation for $\delta P$ is, 
\begin{equation}
\delta P(t) = \int_0^t ds~e^{H(t-s)}~\delta H(s)~P^{lt}(s) \label{eqn:perturbation-solution}
\end{equation}
The change in an observable conditioned to survival is
\begin{equation}
\delta \dfrac{\langle A \rangle_t}{\langle \chi \rangle_t} = \dfrac{\delta \langle A \rangle_t}{\langle \chi \rangle_t^{lt}} - \langle A || \chi \rangle^{lt} \dfrac{\delta \langle \chi \rangle_t}{\langle \chi \rangle_t^{lt}} +  \text{higher order terms}
\end{equation}
which can be obtained from fraction differentiation. We have used the superscript $lt$ to denote that we are starting from the observable average approximately given by Eq. (\ref{eqn:stationary-conditional-average}). Assuming that the perturbation can be factorised, i.e., $\delta H(s) = V~ \delta F(s)$ and $B_{x} \equiv  \frac{1}{\varphi_{1x}} \sum_y V_{x,y} \varphi_{1y}$, the response function is a correlation function. Considering a perturbation affecting a system parameter, we obtain the generalized fluctuation-dissipation theorem similar in form to a standard FDT (see SI Section 2.2 for detailed derivation):
\begin{align}
\delta \dfrac{\langle A \rangle_t}{\langle \chi \rangle_t} =& \int_0^{\infty}ds~\hat{R}_{A,f}(t-s)~\Delta f(s) \\[0.8em]
\hat{R}_{A,f}(t-s) &= -\Theta(t-s) \frac{\partial}{\partial t}\bigg\langle \bigg( A(t) - \chi (t) \langle A || \chi \rangle^{lt} \bigg)~\frac{\partial \hat\phi (s)}{\partial f} || \chi \bigg\rangle^{lt} 
\label{eqn:fdt-discrete-absorbing}
\end{align}
where $\hat\phi$ is now the potential associated with the second right eigenvector, rather than the stationary state, i.e., $\varphi_{1,x} = e^{\hat\phi_{x}}$. The difference between the standard case is immediately apparent from the appearance of the second term which does not exist in Eq. (\ref{eqn:fdt-equilibrium}). In the standard case, this term can be shown to be zero. Consequently, while the standard case and the absorbing boundary case cannot be directly compared, the first term corresponds to the standard FDT with the quasi-stationary distribution used in the place of steady-state distribution. The additional term arises from the fact that the standard FDT considers all states but we need to consider only the survived trajectories, and hence, only the interior states for cases with absorbing states. This term is important as it ensures that the average is performed over trajectories that survive at least until time $t$ which is the time of our observation. 

The response function given in Eq. \eqref{eqn:fdt-discrete-absorbing} is valid for all state-dependent variables, such as mean, variance, skewness, etc. But since Eq. \eqref{eqn:perturbation-solution} is valid for all absorbing processes, we can use it to compute the change in probability distribution dependent quantities such as survival probability and first passage time distribution. The first passage time is defined as the time taken for a system to reach the absorbing state and can range from zero to infinity. Even in the standard theory, an absorbing boundary equivalent state is needed to define these quantities. These state-independent quantities are also significant in many practical examples. 

The survival probability $S(t)$ and the first passage time distribution $\mathcal{P}(t)$ are given by 
\begin{align}
    S(t) &\equiv \sum_{x \neq o} P_x (t) = \langle \chi \rangle_t \\
    \mathcal{P}(t) &= - \partial_t S(t)
\end{align}
Usually, the mean first passage is computed for ease of calculation, but having the full solution for the perturbation and the response function, we can compute the change in the entire first passage time distribution. The resulting response functions for the survival probability and the first passage time distribution due to a perturbation in the parameter $f$ are 
\begin{align}
    &R_{S,f}(t,s) = -e^{-\lambda_1 t} \big[\partial_t \langle \chi(t) \partial_f \phi (s) || \chi \rangle^{lt} \big] \label{eqn:resp-surv} \\
    &\delta \mathcal{P}(t) = -\partial_t \bigg[\int_0^t ds~\Delta f(s)R_{S,f}(t,s) - e^{-\lambda_1 t} \partial_f \lambda_1 \int_0^t ds~\Delta f(s)\bigg] \label{eqn:resp-fpt}
\end{align}

Equations \eqref{eqn:fdt-discrete-absorbing}, \eqref{eqn:resp-surv}, and \eqref{eqn:resp-fpt} are our main results.  Although they are general and strictly valid for all finite and discrete systems, they can be extended to infinite systems that obey the same eigenvalue spectrum as in Eq. \eqref{assum:eigenvalue}. We proceed to show certain applications of these response functions in systems of interest. 

\section{ Applications}
\subsection{Birth-Death Process}

The birth-death process is a common starting point for modelling stochastic population dynamics in ecological communities. A random fluctuation may lead a species to reach zero population, from where it will never recover (without immigration). To investigate how populations approach extinction, we consider a model where the effective (per capita) rates of reproduction and death of an individual are constant. Thus, the rates of the master equation are given by $W_{n+1,n} = b_n = bn$ (birth event), and $W_{n-1,n} = d_n = dn$ (death event). Here the population size of a species is given by the discrete variable $n=0,1,2,\dots$.  Under the neutrality assumption \cite{hubbell2011unified,volkov2003neutral}, the probability distribution of $n$ gives the distribution of the population of all species in the system. When $b<d$ the state $n=0$ is an absorbing boundary. The eventual probability that all species' populations reach zero is one since there are no possible birth events when a species has gone extinct. 

A key point is that the birth-death process is an infinite state system, but still satisfies Eq \eqref{assum:eigenvalue}, because the eigenvalues are discrete and are given by $\lambda_k = (d-b)k$ for $k \geq 1$. Hence, we proceed to use the derived response theory and FDT since the necessary hierarchy is still satisfied.  With $b_{-1}=0$, the Master equation is given by 
\begin{align}
\dot{P}_i(t) = b_{i-1} P_{i-1}(t) + d_{i+1} P_{i+1}(t) - (b_i + d_i)P_i(t) \qquad i \geq 0 \label{bd-me-orig}
\end{align}
The second eigenvector and the eigenvalue can be obtained by the generating function \cite{azaele2016statistical}
\begin{equation}
G(z,t) = \sum_{n\geq 0} P_n(t) z^n =\bigg(\dfrac{1-\mathcal{A}(z,t)}{1 - (1-\nu)\mathcal{A}(z,t)}\bigg)^{n_0} 
\end{equation}
with $P_n(t=0) = \delta_{n,n_0}$, $\nu = 1 - b/d$, $\mathcal{A}(z,t) = (1-z)(1-(1-\nu)z)^{-1}e^{-\nu t}$. At large times $A(z,t)$ is small enough to use the approximation $1/(1-x)^y \approx 1 + yx$. Since the coefficient of the slowest exponential in the long-time state gives the second eigenvector, we obtain $\varphi_{1n} = \nu (1-\nu)^{n-1}$. The time-dependent solution to the Master equation is calculated by using the Meixner Polynomials \cite{karlin1958linear}. (See Supplementary Material Section 3 for the full form of the time dependent solution).

\begin{figure}[H]
	\centering
	\includegraphics[width=14cm]{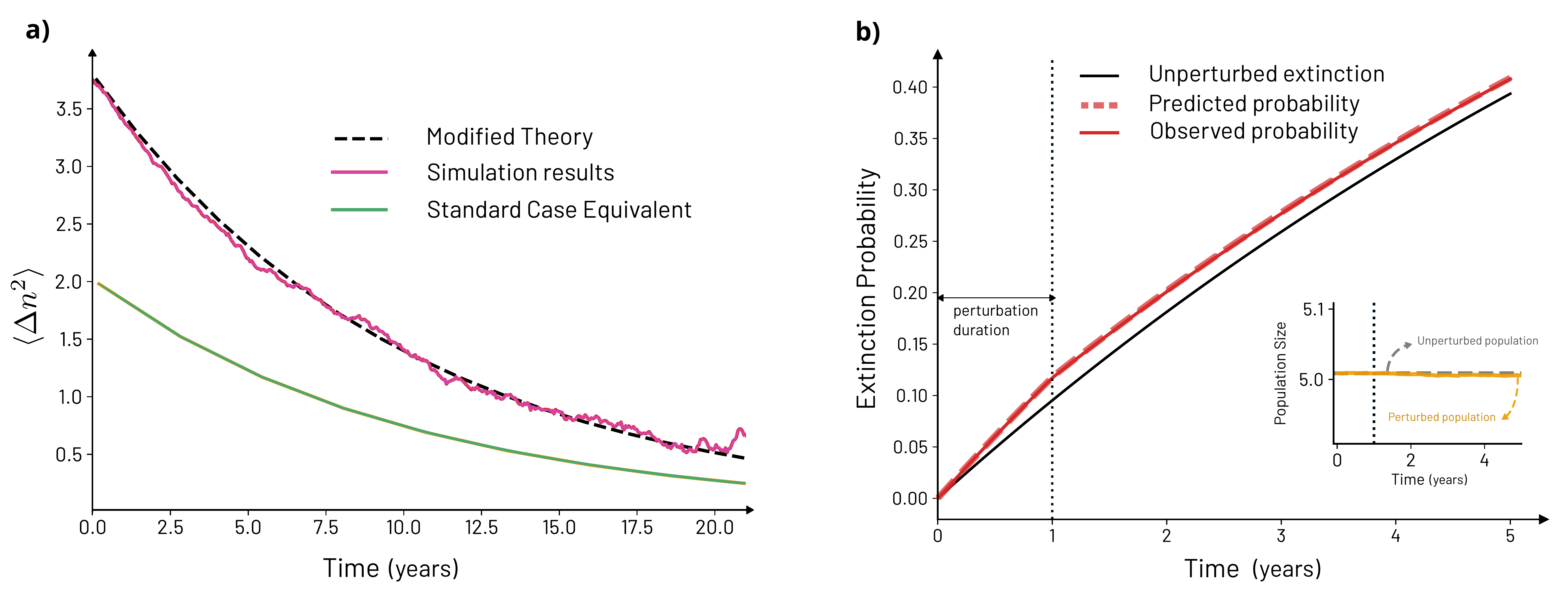}
	\caption{\textbf{Simulated and predicted responses in the birth-and-death process with absorbing boundaries}: \textbf{a)} Comparison between modified response theory Eq. \eqref{eqn:fdt-discrete-absorbing}, standard-equivalent case (similar to Eq. \eqref{eqn:fdt-equilibrium}), and simulations of the system. The observable considered is the average population size squared ($A_n = n^2$). The plots show the deviation of survived observable from the unperturbed value. The green line indicates the contribution from the first term in Eq. \eqref{eqn:fdt-discrete-absorbing}. The black line is the prediction using both terms, and the majenta line is the deviation observed in simulations. The perturbation strength is $\Delta b = b/10$. \textbf{b)} Increase in extinction probability of a species due to sustained perturbation for a period of $t=1$ year in both birth and death rates.  Inset shows the average survived population (from simulations) against time in both perturbed and unperturbed cases. The dotted black line in both plots indicates the time till which perturbation occurs. Solid black line shows the extinction probability in an unperturbed scenario. Dashed lines are predictions, and colored solid lines are observations from simulations with perturbation strengths $\Delta b = b/10$, and $\Delta d =d/10$. Common parameters are $b=0.4$ and $d=0.5$. }
	\label{fig:birth-death}
\end{figure}

Figure \ref{fig:birth-death}a shows the comparison between results from simulation, the new theory given by Eq. \eqref{eqn:fdt-discrete-absorbing}, and the standard-equivalent case given by only the first term of Eq. \eqref{eqn:fdt-discrete-absorbing}. The first term of Eq. \eqref{eqn:fdt-discrete-absorbing} is similar to Eq. \eqref{eqn:fdt-equilibrium}, but with the quasi-stationary distribution in place of the stationary distribution, corresponding to a direct substitution in standard theory. We consider the average population squared, i.e., $\langle n^2 \rangle$, since this shows a significant deviation from standard theory (for details about simulations, see Appendix). It is immediately apparent that the standard-equivalent case significantly underestimates the response of the system while the predictions from the new theory and results from simulations match very well. This indicates that the direct replacement of the distribution is a wrong approach and could result in an incorrect estimate of the response.

An important quantity for ecosystems is the survival probability of species. An increase or decrease in this quantity could mean the difference between extinction and persistence. To study this, we consider a sustained period of perturbation in both rates, reasoning that any increase in death rate caused by natural or man-made causes results in more available resources, thereby also increasing the birth rate. After computing the response function for survival probability, it is possible to compute the extinction probability which is given by $1-\langle \chi \rangle_t$.

Figure \ref{fig:birth-death}b shows the extinction probability for a perturbation in both the birth rate and the death rate  that lasts for a period of $t=1$ year while comparing it to the extinction probability in the unperturbed case. The inset shows the change in average survived population size during and after the perturbation. It is seen that the average population size is unaffected by this perturbation. While not plotted, the quasi-stationary distributions in both cases remain the same, despite the perturbation, which indicates that the observed relative species abundance distribution can appear to be constant in time. The critical point of note is that the extinction probability of the species immediately increases, especially during the period of perturbation, and remains larger than the unperturbed case even after the perturbation has ended. The increase in extinction rate caused by the changing population distribution in both cases is exactly compensated by the changing survival probability, which leads to the constant quasi-stationary distributions, but the increase in death rate is not aptly compensated by the birth rate increase and hence, more species could go extinct. This means that even though the observed average population size and the abundance distribution of the system stay constant, the total number of species in the system decreases faster than usual, and this effect persists beyond the perturbed duration.

While the constant population size and the relative species abundance is ultimately an effect of the particular choice of same perturbation strength for the birth and death rate, the observed phenomenon of increased extinction probability demonstrates the importance of investigating ecological systems using more comprehensive statistical tools.

\subsection{Targeted Search on DNA by Proteins}

We move from the ecological time and length scales to the biochemical scale involving proteins and DNA. Proteins search for a specific set of base pairs on a DNA to which they have a strong binding affinity, thereby forming an absorbing site in the dynamics. We consider a discrete state model of the targeted search for the binding site, which has already been described in the literature \cite{shvets2018mechanisms,iwahara2021discrete}. The model consists of $L-1$ non-specific sites and one specific binding site on the DNA (called `target'). The protein slides along the DNA with a constant diffusion rate $u$ in either direction, until it reaches the target site. At every non-specific site, the protein can get detached from the DNA strand with the rate $k_{off}$ (which determines the average sliding length), and this results in the protein being in free space. Since diffusion in free space is much faster than diffusion in 1D, we consider free space as a single site. From this site, the protein can attach itself to any spot of the DNA strand with the rate $k_{on}$. Due to faster diffusion in 3D, all sites are reachable with equal probability, and hence, $k_{on}$ is the same for all sites. 

Under this framework, we consider rates estimated from observed experimental data (see Appendix), and extract the eigenvalues and eigenvectors of the transition matrix to perform the computations. For all the simulations of the system, the protein starts from free space and moves according to the transitions described above. Eventually, every simulation will end when the protein reaches the target. But, at any given time, a fraction of them will still have the protein not bound to the target site (which we shall term survived realizations, equivalent to survived species in the birth-death process). The long-time probability of finding the protein at a particular site in these survived realizations is equal to the quasi-stationary eigenvector of the transition matrix.

\begin{figure}[h]
    \centering
    \includegraphics[height = 8cm]{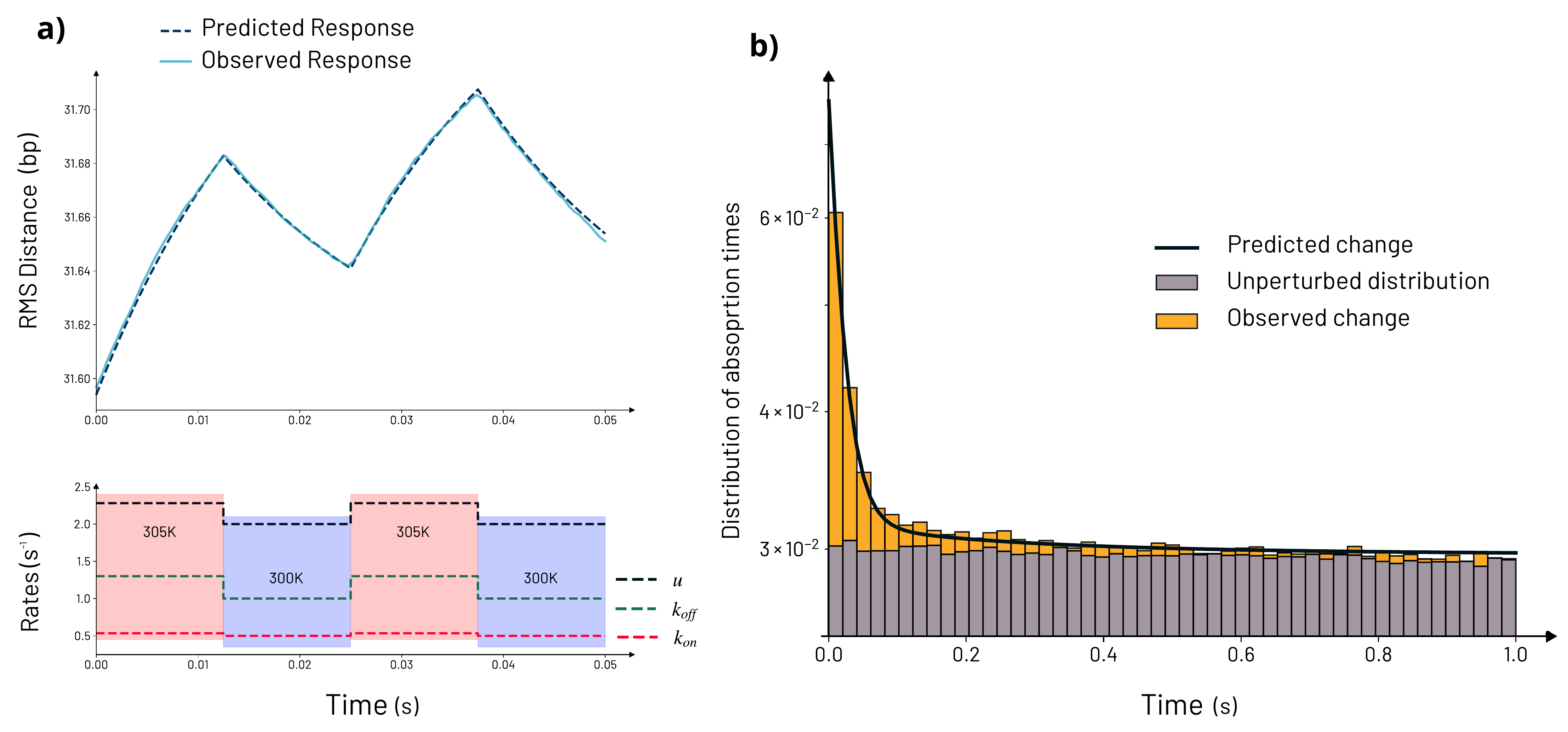}
    \caption{\textbf{Prediction of responses in targeted search by proteins on DNA:} \textbf{a)} Top panel: Root mean squared distance of the protein from the target DNA site for periodic changes in temperature. The RMS distance is given in terms of base pairs, where $1 \text{bp} = 340\text{pm}$ The blue dashed line is the prediction and the solid line is observations from simulations of the protein distance. Bottom panel: Change in the rates of the system assuming Arrhenius rate law. Blue regions indicate the normal temperature (300K), and red regions indicate the higher temperature (305K). The colored dashed lines indicate the rates as given in the legend. \textbf{b)} Prediction of deviation of absorption time distribution with a delta-perturbation of $k_{off}$ in the system. The histogram of absorption time distribution computed through multiple simulations is represented by the bars with the corresponding probability of absorption on the y-axis. Grey bars indicate the unperturbed absorption time distribution, and the orange bars are the deviation upon perturbation.  The solid black line is the prediction from the theory. Perturbation strength is $\Delta k_{off} = k_{off}/10$. The unperturbed first passage time distribution is exponential with a large characteristic time, hence appearing flat at plotted scales. The effect of the delta-perturbation is observed only at times much shorter than the characteristic timescale. See Appendix for the parameter values used.}
    \label{fig:dna}
\end{figure}

Binding and detachment of the protein happen through chemical reactions, whose rates can be approximated by the Arrhenius law, given by $\textit{rate} = \textit{const} \times e^{-E_{\text{rate}}/RT}$, where $E_{\text{rate}}>0$ is the activation energy of the reaction, $R$ is the gas constant, and $T$ is the temperature of the system. Assuming the Arrhenius law for each of the rates, we consider a periodic temperature change of 5$^\circ$C (from 300K to 305K). This results in a perturbation of the rates, which changes the distance of the protein from the target. We compute this change through the root mean squared (rms) distance of the protein from the target. Figure \ref{fig:dna}a shows that even though the observed pattern of change in distance is non-trivial, the new response theory is able to predict it quite well. The specific pattern of whether the protein moves away from the target or towards it depends on the activation energies (see Appendix). In a chosen experimental setting, the activation energies can be first computed and the theory can then be used to predict the change in distance due to temperature change exactly using Eq. \eqref{eqn:fdt-discrete-absorbing} and Eq. \eqref{def:response-function-main}.

The time to extinction is an essential random variable in absorbing processes. The average of this distribution, also called the mean first passage time,  provides important information about the protein reaching the target \cite{iwahara2021discrete}. Figure \ref{fig:dna}b shows the change in the absorption time distribution from simulations and from theory when there is a delta-perturbation in the rate of detachment, $k_{off}$. We see that the probability of absorption at short times immediately increases by a significant proportion. Since a delta-perturbation is considered, at large times, the change in distribution will be minimal, which is seen in the figure where the perturbed and the unperturbed distributions converge at times much larger than the time of perturbation. The flat nature of the unperturbed distribution is due to it being an exponential decay with a characteristic scale much larger than the time which is plotted. The matching between simulations and the theory opens new possibilities in ensemble experiments to control mean first passage time through perturbations.

\section{Conclusions and Discussion}

Our results apply to a broad class of systems that were previously intractable from the lens of response theory. Conditioning to survival is a common tool in mathematical literature of quasi-stationary processes. The generalized fluctuation-dissipation theorem provides insights into how this conditioning affects the response to perturbations. Specifically, this new insight is needed because by direct replacement of the stationary distribution with the quasi-stationary, one significantly underestimates the actual response. Though perturbation of finite linear operators has been studied previously, obtaining the transient dynamics and connections to statistical mechanics and relevant examples are missing to the best of our knowledge. In order to take into account the inherent decay of the system and to predict the response, the generalized Eq. \eqref{eqn:fdt-discrete-absorbing} is necessary.

The two examples we have chosen also demonstrate the wide-ranging applicability of including absorbing states explicitly into the dynamics. These systems are also important from a practical viewpoint. Loss of diversity due to extinctions is a concerning problem. Our results indicate a need for a more robust understanding of decaying ecosystem models by demonstrating that there could be inherent perturbations in the system that increase the extinction probability while leaving often-used population measures intact. This is especially important at small population numbers where fluctuations could drive some species extinct permanently.

At the other end of the size scale, various models have been constructed to account for the variety of interactions that occur in protein search on DNA. With advances in technology, the ability to experimentally verify these models is an exciting new line of research. A potential use of response theory is to design a perturbation to observe a specific response. Taking into account the effect of absorbing states and designing a perturbation to observe a specific first passage time distribution could lead to many potential biomedical applications. Some examples include the release of specific patterns of concentration of drugs, control of resource capture by chemicals or nanoparticles, mediating the response of microbial systems through controlled manipulation of growth media, etc. 

Absorbing processes form a large class of systems which are ubiquitous in nature. Although we have presented the results in the regime of quasi-stationarity, our theory can be easily generalized to systems that have not yet reached quasi-stationarity. This further increases the scope of the work and has the potential to open other doors of investigation, both on the theory and application fronts.

\vspace{2em}

\section*{Appendix}

\textit{Details of Simulations}\\
For the birth-death process, we start all species with an initial population of $n_0 = 100$, using the rates mentioned in the figure captions and simulate until the quasi-stationary regime is reached. Then, a perturbation is applied to the birth rate (which lasts for $dt = 10^{-5}$ in the case of delta perturbation or sustained for a given period that is mentioned). The response of survived species is then tracked over time. This process is repeated for $10^6$ trials to obtain averages.

For the targeted search on DNA, we use the rates mentioned below and simulate the system till time $t = 10/\lambda_1$, where $\lambda_1$ is the leading non-zero eigenvalue and is determined by the transition matrix. For the considered observables in the manuscript, the averages are performed over $10^6$ realizations of survived trajectories. \\

\textit{Rates and activation energies in targeted DNA Search}\\
We consider the diffusion rate, $u = 2 \times 10^5 ~\text{s}^{-1}$, the association rate $k_{on} = 0.5 \times 10^5 ~\text{s}^{-1}$, and the disassociation rate $k_{off} = 1 \times 10^5 ~\text{s}^{-1}$, with $L=100$ sites and the target at site 60. The rates have been approximately determined from figures of experimental data of Egr-1 protein which binds to a specific 8 base-pair sequence of DNA \cite{esadze2014positive,iwahara2021discrete} (rates are chosen at about 400mM concentration of Potassium Chloride (KCl)). To make the calculations easier, we rescale time by dividing all the rates by $10^5$ to get rid of the factor.\\

\noindent \textbf{Code Availability}\\
All the codes used to simulate the examples and the generated data for the plots can be found at \url{https://github.com/prajwalp/responseTheory} \\

\begin{acknowledgments}
P.P. acknowledges the support from the University of Padova through the Ph.D. fellowship within ``Bando Dottorati di Ricerca''. A.M and S.A  authors acknowledge the support of the NBFC to the University of Padova, funded by the Italian Ministry of  University and Research, PNRR, Missione 4 Componente 2, ``Dalla ricerca all’impresa'',  Investimento 1.4, Project  CN00000033. 
\end{acknowledgments}

\end{document}

\maketitle

\section{Fluctuation-Dissipation Theorem in Standard Case}
Consider a system in a steady state or equilibrium. Any external perturbation will shift the system away from this steady state. Assuming that the perturbation is small, the response of the system can be analyzed within the linear approximation. 

Let us consider the Master equation describing our system. The perturbations to the transition matrix can then be written as \cite{risken1996fokker}
\begin{align}
H (t) = H_0 + \delta H (t)
\end{align}

Here, $H_0(x)$ is the unperturbed transition matrix. We consider small perturbations and hence, retain terms upto the first order. Similarly, we also write the solution to the perturbed distribution in terms of the stationary distribution of the master equation and the perturbation:
\begin{align}
P_x(t) = P^{st}_x + \delta P_x(t)
\end{align}

Since we are looking at small perturbations, we also assume that the deviation $\delta P_x(t)$ is small. With $H P^{st} = 0$, we can write:
\begin{align}
\dot P(t) = \dot \delta P(t) = H \delta P(t) + \delta H(t) P^{st} + \delta H(t) \delta P(t)
\end{align}
of which the last term is negligible. Solving this, we obtain 
\begin{align}
\delta P(t) = \bigintsss_{-\infty}^{t}dt'~ e^{H (t-t')} ~\delta H(t') ~ P^{st} 
\end{align}

Having obtained this, we also assume that $\delta H(t)$ can be split into state components and temporal components.
\begin{align}
\delta H_{xy}(t) = \delta H_{xy}~F(t)
\end{align}

Let $A$ be any observable. Then, the deviation of $A$ caused due to the perturbation around its mean value $\langle A \rangle$ can be written as
\begin{align}
&\langle \Delta A(t) \rangle = \int_{-\infty}^{\infty}dt' R_{A,\delta H}(t-t') F(t') \quad \text{where} \\
&R_{A,\delta H}(t) = \Theta (t) \sum_{x,y,z} A_x~[e^{H~t}]_{xy}~\delta H_{yz}~P^{st}_z\label{eq6}
\end{align}

Assuming that the perturbation occurs in one of the parameters of the system, $f \equiv (f_1,f_2 ...,f_N)$, i.e, $f \to  f + \Delta f(t)$. If $H =H(f)$ then,   
\begin{subequations}
\label{eq7}
\begin{align}
& H(f)~P^{st}(f) = 0 \label{eq7a} \\
& P^{st} = e^{\phi(f)} \label{eq7b}
\end{align}
\end{subequations}
\begin{align}
&H(f+\Delta f(t)) = H(f) + \Delta f(t) \delta H + \mathcal{O}(\Delta f^2)\\
&\delta H = \dfrac{\partial H(f)}{\partial f} \label{eq9}
\end{align}

From  \eqref{eq7} and \eqref{eq9}, we can obtain the following 
\begin{align}
\delta H~P^{st}(f) = -H(f)\dfrac{\partial P^{st}(f)}{\partial f}
\end{align}

Subsequently, the response function in \eqref{eq6} can now be written as 
\begin{align}
R_{A,f}(t-t') &= -\Theta(t-t') \dfrac{\partial}{\partial t} \bigg \langle A(t)~\dfrac{\partial \phi}{\partial f}(t') \bigg \rangle \nonumber \\[2em]
& = -\Theta(t) \frac{d}{dt} \bigg\langle A(t) \dfrac{\partial \phi(t'=0)}{\partial f} \bigg\rangle \label{eq14}
\end{align}

Suppose we denote $X_f = \dfrac{\partial \phi(t=0)}{\partial f}$, then, (\ref{eq14}) can be written as 
\begin{align}
R_{A,X_f}(t) = -\Theta(t) \frac{d}{dt} K_{A,X_f}(t) \label{eqFDT}
\end{align}
where $K_{A,B} (t)$ is the two time correlation function between observables $A$ and $B$. Equation \eqref{eqFDT} forms the main Fluctuation Dissipation Theorem we wish to pursue. We are considering the perturbation happening to a stationary distribution. Though we have presented the discrete case here, the same calculations hold for continuous state space with the Fokker-Planck operator being analogous to the transition matrix $H$ \cite{risken1996fokker}.

\section{Discrete System with Absorbing Boundaries}

\subsection{Master Equation and Preliminaries}
Let $\Omega$ represent the space of all the states. We can denote the boundary of this space by $\partial \Omega$ and the interior by $\Omega^{\circ} = \Omega - \partial \Omega$. The transition rates between different states shall be represented by $W_{y \rightarrow x} = W_{xy}$. The absorbing boundary condition can be represented by setting the rate of transition out of the boundary states to zero.
\begin{align*}
W_{xy} = 0 &\quad \forall x \in \Omega, y \in \partial \Omega \\
W_{xy} \geq 0 &\quad
\end{align*}

The master equation for the time evolution of the probability of being in state $x$ at time $t$ is
\begin{align}
\dot{P}_x(t) &= \sum_{y \in \Omega} \big[ W_{xy}~P_y(t) - W_{yx}~P_x(t)\big] \nonumber \\
&= \sum_{y \in \Omega^{\circ}} W_{xy}P_y(t)  &\text{if $x \in \partial \Omega$}  \nonumber \\
\text{or} \quad &= \sum_{y \in \Omega^{\circ}} W_{xy}P_y(t) - \sum_{y \in \Omega } W_{yx}P_x(t)  &\text{if $x \in \Omega^{\circ}$} 
\end{align}

Let a single state $0$ represent all the absorbing states. The corresponding entries for this state that would go into the ME would be
\begin{align}
&P_0(t) \equiv \sum_{x \in \partial \Omega} P_x(t) \label{prob-redef-0} \\ 
&W_{0x} \equiv \sum_{y \in \partial \Omega} W_{yx} \quad \text{and} \quad W_{x0} \equiv \sum_{y \in \partial \Omega} W_{xy} = 0 \label{rate-redef-0}
\end{align}
Note that these also imply $W_{0x} = 0$ if $x \in \partial \Omega$.

With new state space $\Omega' = \Omega^{\circ} \cup \{0\}$, the ME is
\begin{equation}
\dot{P}_x(t) = \sum_{y \in \Omega'} \big[ W_{xy}~P_y(t) - W_{yx}~P_x(t) \big]
\end{equation}

\subsubsection{Eigenvales and Eigenvectors}

$\sum_x \dot{P}_x(t) = 0$, i.e, the total probability is conserved (when absorbing states are also considered). We can define the operator $H$ with which we have $\dot{P}(t) = H~P(t)$
\begin{equation}
H_{xy} = W_{xy} - \delta_{xy} \sum_z W_{zy}
\end{equation}

Normalizing the left and right eigenvectors of $H$ such that $\langle \psi^n | \varphi_m \rangle = \delta_{m,n}$, the first left eigenvector is $\psi^0_x=1 \quad \forall x \in \Omega$ for the eigenvalue $\lambda_0 = 0$. The corresponding right eigenvector is $\varphi_{0x} = \delta_{x,0}$.

In general, for the $n^{th}$ right eigenvector and eigenvalue, with $\Re(z)$ representing the real part of a complex number $z$,
\begin{align}
&H\varphi_n = -\lambda_n  \varphi_n \nonumber \\ 
\text{and hence} \quad& 0 = \lambda_0 < \Re(\lambda_1) < \Re (\lambda_2) \leq \Re (\lambda_3) ... \label{eigenvalue-assumption}
\end{align}
We will always work with an irreducible $W$, i.e. for any pair of interior nodes, there is always a path of non zero $W$'s from one to the other. This implies that the boundary state is not visited. In such a system, \eqref{eigenvalue-assumption} can be proved using Perron-Frobenius Theorem \cite{seneta2006non}. Note that the inequality between $\lambda_1$ and $\lambda_2$ is different from the one between the higher eigenvalues. This is due to the spectrum possibly being complex and hence, having complex conjugate eigenvalues which have equal real parts. Due to the Perron-Frobenius theorem, $\lambda_1$ will always be real and hence, have no conjugate counterpart. \\

\noindent \textbf{Eigenvalue Spectrum}

Consider an infinitesimal increment in time $\epsilon$. Then, the Master equation can be discretized to the Markov chain (at small $\epsilon$)
\begin{align}
P(t+\epsilon) = P(t) + \epsilon H~P(t)
\end{align}

Consequently, define $T \equiv (\mathbb{I} + \epsilon H)$ where $\mathbb{I}$ is the identity matrix. $T$ is a non negative, irreducible matrix of size $ N \times N$ where $N$ is the number of states. The eigenvalues of $T$ are simply $\alpha_i = 1 - \epsilon \lambda_i$. Consider the submatrix of $T$ corresponding to the interior, $T^*$. Due to irreducible system, there exists a path from one state to another, and hence, there always exists a $k_{xy}>0$ for which the $(W_{xy})^{k_{xy}}$ is non zero. Choosing the largest of these $(N-1)^2$ positive numbers ensures that $T^*$ is primitive, i.e, $(T^*)^{k_{largest}} >0$, if $\epsilon$  is sufficiently small.

The characteristic function of the eigenvalues of $T^*$ is the characteristic function of $T$ sans the $(\alpha-1)$ factor. Hence, the eigenvalues of $T^*$ is equivalent to the eigenvalues of $T$. Note that $T^*_{xy} \geq 0$ when $x \neq y$ and $T^*_{xx} = 1 - \epsilon \sum_z W_{zx}$ which is bounded allowing us to choose sufficiently small $\epsilon$ such that $T^{*}_{xy} \geq 0$ $\forall x,y$. Thus, we can  apply the  Perron–Frobenius theorem \cite{seneta2006non} on the matrix $T^*$. Then,the leading real eigenvalue $\alpha_1 > 0$. For any other eigenvalue $\alpha_2$, then, $\alpha_1 > | \alpha_2|$. Then, with $\Re(z)$ and $\Im(z)$ representing the real and imaginary parts of $z$,
\begin{align}
&\alpha_1 > |\alpha_2| \nonumber \\
\implies & ( 1 - \epsilon \lambda_1)^2 > | 1 - \epsilon \Re(\lambda_2) - i \epsilon \Im(\lambda_2) |^2 \nonumber  \\
\implies & \epsilon^2 \lambda_1^2 - 2 \epsilon \lambda_1 > \epsilon^2 (\Re(\lambda_2)^2 + \Im(\lambda_2)^2) - 2 \epsilon \Re(\lambda_2)
\end{align}
In the limit $\epsilon \to 0$, which is the correct limit to get the Master equation back, we have $\Re(\lambda_2)> \lambda_1$. 

The Perron-Frobenius Theorem also guarantees that the left and the right eigenvector of $T^*$ corresponding to the eigenvalue $\alpha_1$ (equivalently, $\lambda_1$) are non-degenerate and have components strictly positive. 

If $\varphi_1$ and $\psi^1$ are the right and left eigenvectors of $H$ corresponding to $-\lambda_1$, then, $\varphi_{1x},\psi^1_{x} > 0$ $\forall x \neq 0$. Hence, we can choose the normalization such that $\sum_{x \neq 0} \varphi_{1x}  = 1$. Since $0 = \sum_{x} \psi^0_x \varphi_{1x} = \sum_x \varphi_{1x} \implies \varphi_{10} = -1$. Then, 
\begin{align}
\lambda_1 = -\lambda_1 \varphi_{10} = \sum_{y}H_{0y}~\varphi_{1y} = \sum_{y\neq 0} W_{0y}~\varphi_{1y} > 0
\end{align}
if at least one of the $W_{0y}$ is positive. 

\subsubsection{Average and Correlation Functions}

Consider a specific initial condition $P^{lt}(0) = \varphi_0 + \varphi_1$. Then, the evolution starting from this initial condition is given by $P^{lt}_x (t)  = \varphi_{0x} + e^{-\lambda_1 t} \varphi_{1x}$. Defining an observable $\chi_x = 1 - \delta_{x,0}$, whose average value at time $t$ gives the survival probability, i.e. the probability that the system has not yet been absorbed by the boundary. For any observable $A$, we consider it to be zero on the boundary since we wish to know only survived averages. Any observable $B$ which is non-zero on the boundary can be redefined as $B\chi$ such that it is zero on the boundary states.

\begin{align}
&\langle A \rangle_t^{lt} = \sum_x A_x P^{lt}_x(t) = e^{-\lambda_1t} \sum_{x\neq 0} A_x \varphi_{1x} \nonumber \\
&\langle \chi \rangle_t^{lt} = e^{-\lambda_1t} \sum_{x \neq 0} \varphi_{1x} = e^{-\lambda_1 t} \qquad \text{this represents the survival probability} \nonumber \\
&\text{Conditional Average} \nonumber \\
&\dfrac{\langle A \rangle_t^{lt}}{\langle \chi \rangle_t^{lt}} = \sum_{x \neq 0} A_x \varphi_{1x} \equiv \langle A  || \chi \rangle^{lt} \label{conditional-one-operator}
\end{align}

For an arbitrary initial condition (IC), at large times, it is equivalent to averaging over the defined $P^{lt}$ as seen from below. For arbitrary $P(0)$,
\begin{align*}
&P(t) = \varphi_0 + c_1 e^{-\lambda_1 t} \big[ \varphi_1 + \mathcal{O}(e^{-t (\Re(\lambda_2) - \lambda_1)})\big] \nonumber \\
&\langle A \rangle_t = \sum_x A_x P_x(t) = c_1 e^{-\lambda_1t} \sum_{x\neq 0} A_x \varphi_{1x}\big[ 1 + \mathcal{O}(e^{-t (\Re(\lambda_2) - \lambda_1)})\big] \nonumber \\
&\langle \chi \rangle_t = c_1 e^{-\lambda_1t} \sum_{x \neq 0} \varphi_{1x} \big[ 1 + \mathcal{O}(e^{-t (\Re(\lambda_2) - \lambda_1)})\big] = c_1 e^{-\lambda_1 t} \big[ 1 + \mathcal{O}(e^{-t (\Re(\lambda_2) - \lambda_1)})\big] \nonumber \\
&\dfrac{\langle A \rangle_t}{\langle \chi \rangle_t} =_{t \rightarrow \infty} \sum_{x \neq 0} A_x \varphi_{1x} = \langle A  || \chi \rangle^{lt}
\end{align*}
This gives us the conditional average. For the time correlation function, we  consider the propagator. 

\subsubsection*{Time Evolution}
The ME can be written as matrix operation with the $H$ defined earlier. 
\begin{align*}
&\dot{P}(t) = H ~P(t) \quad \implies  P(t) = e^{H t} P(0)
\end{align*}
If $P_x(0) = \delta_{x,x_0}$, we obtain the propagator $P(x,t|x_0,0)=[e^{Ht}]_{x,x_0}$. 
Hence, for a general IC, 
$$
P_x(t) = \sum_{x_0} P(x,t|x_0,0) P_{x_0}(0)
$$
Since $H_{x,0}=0$ we have
\begin{equation}\label{propx0}
    P(x,t|x_0=0,0)=[e^{Ht}]_{x,0}=\delta_{x,0} \qquad \forall t
\end{equation}

Correlation functions are defined as $\langle A(t) B(t') \rangle = \sum_{x,y} A_x ~P(x,t|y,t')~B_y P_y(t')$. 
It follows from \eqref{propx0} and the Chapman Kolmogorov equation \cite{van1992stochastic} that 
\begin{equation}
    \langle A(t) B(t') \cdots C(t'') \chi (t''')\rangle = \langle A(t) B(t')\cdots C(t'')\rangle \quad \text{if} ~ t'''\leq \max\{t,~t'\dots t''\}
\end{equation}
This has to be true logically too. Since we are calculating the correlation till $\max \{t,t'\dots t''\}$, the only contributing trajectories have to last at least till time $\max\{t,~t'\dots t''\}$. Hence, at time $\leq t'''$ the trajectories have not yet hit the boundary, i.e. $\chi (t''')=1$. But for the contrary case, this is not true
\begin{equation}
    \langle A(t) B(t') \cdots C(t'') \chi (t''')\rangle \neq \langle A(t) B(t')\cdots C(t'')\rangle \quad \text{if} ~ t'''> \max\{t,~t'\dots t''\}
    \label{multiple-time-survived}
\end{equation}
since on the l.h.s. we are requiring that trajectories have not hit the boundary at least till time $t'''$ whereas in the r.h.s.  we require that they only survive till a time $ \geq t''$. This is a consequence of having absorbing boundary. In the standard no absorbing boundary case, $\chi = 1$ trivially and $\sum_{x}P(x,t|y,t') = 1$. With the absorbing boundary, we are forced to neglect the $0$ state and hence the sum over $x \neq 0$ is no longer unity. 
If the IC is $P^{lt}$,
\begin{align}
\langle A(t) B(t') \rangle^{lt} &= \sum_{x,y} A_x~ B_y ~P(x,t|y,t')~P_y^{lt}(t') \nonumber \\
&= \sum_{x,y} A_x~B_y~P(x,t|y,t')~\varphi_{1y}~e^{-\lambda_1 t'} \qquad t'<t
\end{align}
In this case, a conditional correlation function, which represents the correlation with only surviving trajectories is written as 
\begin{align}
\dfrac{\langle A(t) B(t') \rangle^{lt}}{\langle \chi \rangle_t^{lt}} &= \sum_{x,y} A_x~B_y~P(x,t|y,t')~\varphi_{1y}~e^{\lambda_1 (t-t')} \nonumber \\
&= Tr[A~e^{(H+\lambda_1 \mathbb{I})(t-t')}~B~\Phi_1] \equiv \langle A(t)~B(t') || \chi \rangle^{lt} \qquad t'<t \label{two-point-trace}
\end{align}
where $\mathbb{I}$ is the identity matrix. In this, we define $A_{xy} = A_x \delta_{xy}$ and $B_{xy} = B_y \delta_{xy}$. $(\Phi_1)_{xy} = \varphi_{1x}~\delta_{xy}$. Similar to the conditional average, we can start from arbitrary IC and at large times, it becomes equivalent to averaging over the long time state. Hence, in further calculations, we can just use $P^{lt}(t)$ to determine correlations at large times. \footnote{In practice in order to calculate conditional averages at large times, as defined above (see for example \eqref{two-point-trace}), we can use the following formula
\begin{equation}
    \langle A(t)~B(t') || \chi \rangle^{lt}=\lim_{t_0\rightarrow -\infty}\frac{\sum_{x,y,z} A_x~P(x,t|y,t')~B_y~P(y,t'|z,t_0)~P_{initial}(z)}{\sum_{x,z} \chi_x~P(x,t|z,t_0)~P_{initial}(z)}
\end{equation}
and similarly for multiple time correlation functions. The initial condition $P_{initial}$ at time $t_0$ is arbitrary as far as $P_{initial}(0)<1$. But for our purposes, it is easier to choose $P_{initial}(0)=0$.}

At large time limit, $t' \rightarrow \infty$ and denoting  $\Delta t = t-t'$, the propagator can be approximated as
$$
\big[e^{H \Delta t}]_{xy} = \psi^0_y \varphi_{0,x} +  \psi^1_y\varphi_{1x} e^{-\lambda_1 \Delta t}+ \mathcal{O}(e^{-\Re(\lambda_2)\Delta t})
$$
Using this in \eqref{two-point-trace}, 
\begin{equation}
\langle A(t)~B(t') || \chi \rangle^{lt} = \sum_{x}A_x~\varphi_{1x} ~\sum_y B_y ~\varphi_{1y}~ \psi^1_y \big( 1 + \mathcal{O}(e^{-(\Re(\lambda_2) - \lambda_1)\Delta t}\big)
\end{equation}
By defining $\langle B || \chi \rangle^{lt}_{cond} \equiv \sum_y B_y~\varphi_{1y}~ \psi^1_y = \langle B~\psi^1 || \chi \rangle^{lt}$. Therefore, at large time limit, $t-t'\rightarrow \infty$, 
\begin{equation}
\langle A(t)~B(t') || \chi \rangle^{lt} = \langle A || \chi \rangle^{lt} ~ \langle B || \chi \rangle^{lt}_{cond} ~ \big( 1 + \mathcal{O}(e^{-(\Re(\lambda_2) - \lambda_1)\Delta t}\big) \label{two-point-cond-defn}
\end{equation}
Since $\psi^1_x > 0$, we also have $\langle B || \chi\rangle^{lt}_{cond} >0$ if $B_x \geq 0$. If $B = \chi$, $\langle \chi || \chi \rangle^{lt}_{cond} = \sum_{y \neq 0} \varphi_{1y}\psi^1_y = \sum_y \varphi_{1y}~\psi^1_y = 1$ (since $\psi^1_0 = 0$ thanks to the normalization of left and right eigenvectors). It is important to note that this is a significant difference from the standard treatment where $\langle A(t)B(t') \rangle^{st} = \langle A \rangle^{st} \langle B \rangle^{st}$ at large time separation. This is because in the standard case with a non trivial stationary state, $\varphi_0$, one has $\psi^0_{x}=1$ for all $x$.  This can also be generalized to multiple correlations 
\begin{align}
\langle A(t)~B(t')~\cdots C(t'') || \chi \rangle^{lt} =& \langle A || \chi \rangle^{lt} ~ \langle B || \chi \rangle^{lt}_{cond}~\cdots \langle C || \chi \rangle^{lt}_{cond}  \times \nonumber \\[0.9em]
& \times \big( 1 + \mathcal{O}(e^{-(\Re(\lambda_2) - \lambda_1)\min(\Delta t,\Delta t'~\cdots)}\big)
\label{multiple-point-cond-defn}
\end{align}
when $t''< t'<t$ in the $t-t'=\Delta t$ and $t'-t''=\Delta t'$ large limit. 

Using some examples illustrates this point and provides further insight. If we have $A=\chi$, then, $\langle \chi (t) B(t') || \chi \rangle^{lt} \rightarrow \langle B || \chi \rangle^{lt}_{cond}$. This is in contrast to the standard treatment. Correlation in standard treatment would be definied using the stationary state, i.e, the non trivial $P_{st} = \varphi_0$. Standard correlation, $K_{A,B}^{st} = \sum_{x,y} A_x~P(x,t|y,t')~B_y~P_{st}(y)$. Therefore, if $A_x = 1$, $K_{1,B}^{st} = \sum_{x,y} P(x,t|y,t')~B_y~P_{st}(y) = \sum_y B_y~P_{st}(y) = \langle B \rangle_{st}$

\textbf{Remark.} Looking at the previous two examples, we can see that using $A=\chi$ the average is over trajectories that survive in the interior and don't get absorbed by the boundary at least till time $t$ where $A=\chi$ is evaluated. So, we are essentially averaging over only the remaining/surviving particles/trajectories. 

With these preliminaries in place, we can now look at how a perturbation in the system affects the observables. We now proceed to look at linear response to perturbation in such systems. 

\subsection{Linear Response}

To capture the effect of perturbation, let us consider the perturbation to be represented by a change in the rates, i.e, consequently a change in $H$. The effect of this perturbation will be a change in the long time state. 
\begin{align*}
&H \rightarrow H + \delta H(t) \\
& P^{lt}(t) \rightarrow P^{lt}(t) + \delta P(t)
\end{align*}
where $P^{lt}(t) = \varphi_0 + c_1 e^{-\lambda_1 t}\varphi_1$ which is a more general version of the long time state starting from $P^{lt}(0)$ defined in the earlier subsection.
Therefore, we can write an equation for the time evolution of $\delta P(t)$ and solve it as a first order ODE.
\begin{align}
&\dot{P}^{lt}(t) + \dot{\delta {P}}(t) = \big(H+\delta H(t)\big)\big(P^{lt}(t) + \delta P(t)\big) \\
&\dot{\delta {P}}(t) = H~\delta P(t) + \delta H(t) ~ P^{lt}(t) +  \text{higher order terms}
\end{align}
The solution of this with initial condition, $\delta P(0) = 0$ is
\begin{equation}
\delta P(t) = \int_0^tds~e^{H(t-s)}~\delta H(s)~P^{lt}(s)
\end{equation}
Then, the change in our conditional average of observable is going to be
\begin{align}
&\delta \dfrac{\langle A \rangle_t}{\langle \chi \rangle_t} = \dfrac{\delta \langle A \rangle_t}{\langle \chi \rangle_t^{lt}} - \langle A || \chi \rangle^{lt} \dfrac{\delta \langle \chi \rangle_t}{\langle \chi \rangle_t^{lt}} +  \text{higher order terms}
\end{align}

\begin{align*}
\dfrac{\delta \langle A \rangle_t}{\langle \chi \rangle_t^{lt}} &= \sum_x A_x\delta P_x(t) / \langle \chi \rangle_t^{lt} \\
&=\frac{1}{\langle \chi \rangle_t^{lt}} \int_0^t ds \sum_{x,x',y} A_x~P(x,t|x',s)~\delta H_{x',y}~P^{lt}_y(s)
\end{align*}

Let us assume that the perturbation can be split into time and state components, i.e, $\delta H(s) = V~\delta F(s)$, with $V_{x,0}=0$. Then, we can write the perturbation of the operator average as a response function times the time component of the perturbation. That is,
\begin{equation}
\dfrac{\delta \langle A \rangle_t}{\langle \chi \rangle_t^{lt}} = \int_0^{\infty} ds~R_{A,V}(t,s)~\delta F(s)
\end{equation}
with $R_{A,V}(t,s) = \dfrac{\Theta(t-s)}{\langle \chi \rangle_t^{lt}} \sum_{x,x',y} A_x~P(x,t|x',s)~V_{x',y}~P^{lt}_y(s) \equiv \Theta(t-s)~\langle A(t)~B(s) || \chi \rangle^{lt}$\\[1em] and
\begin{align}
B_{x'} =  \frac{1}{\varphi_{1x'}} \sum_y V_{x',y} \varphi_{1y} \label{b-v-relation}
\end{align}

Similarly we also write $\dfrac{\delta \langle \chi \rangle_t}{\langle \chi \rangle_t^{lt}} = \int_0^{\infty}ds~R_{\chi,V}(t,s)~\delta F(s)$. Ultimately, with the defined $B$, we have
\begin{align}
&\delta \dfrac{\langle A \rangle_t}{\langle \chi \rangle_t} = \int_0^{\infty}ds~\hat{R}_{A,V}(t-s)~\delta F(s) \\[0.8em]
&\hat{R}_{A,V}(t-s) = \Theta (t-s) \big[ \langle A(t)~B(s)|| \chi \rangle^{lt} - \langle A || \chi \rangle^{lt}~\langle \chi (t)~B(s) || \chi \rangle^{lt} \big] \label{final-response-form}
\end{align}
Notice that, thanks to \eqref{two-point-cond-defn} and $\langle \chi || \chi \rangle^{lt} = 1$, we have $\lim_{t\rightarrow\infty}\hat{R}_{A,V}(t)=0$. Now that we have the form of the response function with conditional correlations, we can try and understand the fluctuation dissipation theorems associated with it.

\subsubsection*{Fluctuation Dissipation Theorem}
Similar to the standard case, let us assume that $H$ depends on some parameter $f$ (can also be a set of parameters. For simplicity, we consider one, but the calculations remain the same for multiple parameters). We can now Taylor expand $H$ around no perturbation in $f$
\begin{align*}
&H(f + \Delta f) = H(f) + \dfrac{\partial H(f)}{\partial f}\bigg|_{\Delta f = 0}~\Delta f + \mathcal{O}(\Delta f^2)
\end{align*}
We drop the subscript $\Delta f = 0$ for visual simplicity. We see that $V = \partial_f H(f)$. Since $(H+\lambda_1 \mathbb{I})\varphi_1 = 0$, differentiating, 
\begin{equation}
\bigg( V + \dfrac{\partial \lambda_1}{\partial f} \mathbb{I} \bigg)~\varphi_1(f) + \big( H(f) + \lambda_1 \mathbb{I} \big)~\dfrac{\partial \varphi_1}{\partial f}(f) = 0
\end{equation}

Using the above and  \eqref{two-point-trace} and \eqref{b-v-relation}, we can write the correlation function 

\begin{align}
\langle A(t)~B(s) || \chi \rangle^{lt} &= \sum_{xy} A_x \bigg( e^{(H(f) + \lambda_1 \mathbb{I} )(t-s)} \bigg)_{xy}~(V \varphi_1)_y \nonumber \\[1em]
&= \bigg[ -\dfrac{\partial}{\partial t} \langle A(t)~\frac{\partial \phi (s)}{\partial f} || \chi \rangle^{lt} - \langle A|| \chi \rangle^{lt} \frac{\partial \lambda_1(f)}{\partial f} \bigg] \label{simlpified-conditional-with-pote}
\end{align}

In \eqref{simlpified-conditional-with-pote}, if we use $A=\chi$, we get a similar form for the second term in \eqref{final-response-form} and we finally obtain (we have dropped the $f$ dependence in $\phi$ only for simplicity of expression)
\begin{equation}
\hat{R}_{A,V}(t-s) = -\Theta(t-s) \frac{\partial}{\partial t}\bigg\langle \bigg( A(t) - \chi (t) \langle A || \chi \rangle^{lt} \bigg)~\frac{\partial \phi (s)}{\partial f} || \chi \bigg\rangle^{lt} \label{fdt-final-form}
\end{equation}

Equation (\ref{fdt-final-form}) is the \textbf{Fluctuation Dissipation Theorem} for our case of absorbing state in finite and discrete systems! Notice that
\begin{equation}
\bigg\langle \bigg( A(t) - \chi (t) \langle A || \chi \rangle^{lt} \bigg)|| \chi \bigg\rangle^{lt}=0.
\end{equation}
In the standard treatment, the second term involving $\chi(t)\langle A ||\chi \rangle$ in \eqref{fdt-final-form}, is absent. Its presence is important since, from, the remark at the end of the previous section, we expect that, in general
\begin{equation}\label{last-term}
    \frac{\partial}{\partial t}\bigg\langle \chi (t)~\frac{\partial \phi (s)}{\partial f} || \chi \bigg\rangle^{lt}\neq 0
\end{equation}

\section{Birth Death Process}
The eigenvalues for the Birth Death Process are $\lambda_i = (d-b)i$ \cite{karlin1958linear} which can also be deduced from $G(z,t)$. We also have the solution for the Master Equation \cite{karlin1958linear}. A point of note is that we differ in the absorbing state used in \cite{karlin1958linear}. While -1 is an absorbing state in \cite{karlin1958linear}, we choose 0 as the absorbing state. Therefore, the solution to the Master equation will be same as in \cite{karlin1958linear} but with the index values shifted by 1 unit.  We represent $P_{n,n_0}(t)$ as probability of being at $n^{th}$ state at time $t$ starting from state $n_0$ at time $t=0$. We define

\begin{align}
q_{n_0,n}(t) &= \bigg(1-\frac{b}{d} \bigg)^2 \bigg(\frac{b}{d}\bigg)^{n-1} \frac{n_0}{n} (1-e^{-\lambda_1 t})^{n_0-1} \bigg(1-\frac{b}{d} e^{-\lambda_1 t}\bigg)^{-n_0-1} \nonumber \\
& \times \sum_{k=0}^{n_0-1} \binom{n_0-1}{k} (-1)^k \bigg[ \dfrac{1 - \frac{d}{b} e^{-\lambda_1 t}}{ 1 - e^{-\lambda_1 t}}\bigg]^k \bigg[\dfrac{ 1 - e^{-\lambda_1 t}} {1 - \frac{b}{d} e^{-\lambda_1 t}}\bigg]^{n-1-k} \dfrac{(n_0 + n + k)!}{(n-1-k)! ~n_0 !} \label{bd-q-soln}
\end{align}

The summation over $k$ can be written in a compact form as 
$$
\frac{\Gamma (n_0+n) \left(\frac{d \left(e^{-\lambda_1 t}-1\right()}{b e^{-\lambda_1 t}-d}\right)^{n-1} \, _2F_1\left(-n_0-1,-n-1;-n_0-n-3;\frac{\left(d e^{b t}-b e^{d t}\right) \left(b e^{b t}-d e^{d t}\right)}{b d \left(e^{b t}-e^{d t}\right)^2}\right)}{\Gamma (n_0+1) \Gamma (n)}
$$
where $_2F_1$ is the hypergeometric function. In terms of the above definition we have:
\begin{align}
P_{n,n_0}(t) &= \bigg(\frac{b}{d}\bigg)^{n-n_0}~ \dfrac{n_0}{n}~ q_{n,n_0}(t) \qquad \text{if} \quad n_0 >n \nonumber \\ 
&= q_{n_0,n}(t) \qquad \text{otherwise} \label{bd-solm}
\end{align}

\noindent \textbf{Response Function}

Since we are dealing with an infinite dimensional operator, we do not know how $e^{H t}$ acts on $\partial_f~ \varphi_1$. Hence, we rewrite (\ref{fdt-final-form}) slightly.  
\begin{align}
R_{A,f}(t) = - \frac{\partial}{\partial t} \bigg( \bigg\langle  A(t)\frac{\partial \phi (0)}{\partial f} || \chi \bigg\rangle -  \langle A || \chi \rangle \bigg\langle \chi (t)~\frac{\partial \phi (0)}{\partial f} || \chi \bigg\rangle \bigg) 
\end{align}

\begin{align}
\bigg\langle  A(t)\frac{\partial \phi (s)}{\partial f} || \chi \bigg\rangle &=  \sum_{x,y} A_x \bigg(e^{Ht)}\bigg)_{x,y}e^{\lambda_1 (t-s)} \frac{\partial \phi_{1y}(s)}{\partial f} \varphi_{1y}
\end{align}

Since $P(t) = e^{Ht}P(0)$, with $P_n(0) = \delta_{n,n_0}$, we have $ (e^{Ht})_{x,y} = P_{x,y}(t)$. 
\begin{align}
\frac{\partial}{\partial t} \bigg\langle  A(t)\frac{\partial \phi (s=0)}{\partial f} || \chi \bigg\rangle &= \frac{\partial}{\partial t}  \sum_{x,y} A_x P_{x,y}(t)e^{\lambda_1 t} \frac{\partial \phi_{1y}}{\partial f} \varphi_{1y} \nonumber \\[0.9em]
&= \sum_{x,y} A_x \big( \lambda_1 P_{x,y}(t) + \dot{P}_{x,y}(t) \big) e^{\lambda_1 t} \frac{\partial \phi_{1y}}{\partial f} \varphi_{1y}
\end{align}

where we can use master equation for $\dot{P}_{x,y}(t)$. While closed form expression for $R_{A,f}$ is difficult to calculate, we can compute it approximately using the explicit formula for $P_{x,y}(t)$ from Eq. (\ref{bd-solm}) and (\ref{bd-q-soln}) . 

\newpage

\bibliographystyle{plainnat}